\documentstyle[12pt,qqaalart]{article}

\author{Manuel Casta\~ns\thanks{E-mail: miguel@iponet.es}\\
Deprt. F\`isica\\
ETS Arquitectura\\
UPM Madrid \\
Juan de Herrera 4 \\
Spain
\and
J.A. Belinch\'on\thanks{E-mail: jabelinchon@usa.net}\\
Dpt. F\`isica\\
ETS Arquitectura\\
UPM Madrid\\
Juan de Herrera 4\\
Spain}
\title{Enlargement of Planck System of Absolute Units
}
\date{ 
}
\input tcilatex

\QQQ{Language}{
American English
}

\begin{document}

\maketitle
\begin{abstract}
Conventional Planck system of absolute units $(G,c,\hbar ,k_B)$ is enlarged
by the addition of the so-called, dynamical constant $\Gamma $ , that
relates force to the rate of change of linear momentum. A further
enlargement allows its applications to electromagnetism by introducing the
permitivity of the vacuum $\epsilon _0$, as a new absolute unit. Consequenly
there are six reference dimension $(M,L,T,F,\Theta ,I)$ where $F$ and $I$
stand for dimensions of force and electric current respectively. In such an
'' enlarged Planck system '' ( E.P.S) we equalize its unit of charge to the
elementary one, $e$, with the result that $\Gamma $ coincide with the
fine-structure constant, $\alpha .$ In the proposed E.P.S. the constants $%
G,c,\hbar ,k_B,\epsilon _0,\mu _0,\Gamma ,e,\alpha $ acquire very simple
numerical values, (most of them are equal to the unity), although in our
view they retain their physical dimensions. A calculation of the main
quantities in the E.P.S. and some considerations that may concern
homogeneity and isotropy conditions in the very early universe are made.
\end{abstract}

\section{\bf Planck System.}

Admitedly the Planck absolute system of units ( P.A.S.U.) rests on the
gravitational constant $G$ , quantum of action $\hbar $, velocity of light $%
c $ and Boltzmann`s constant $k_B$ . The units of length (Planck length $l_p$%
), of time ($t_p$), of mass ($m_p$) and the temperature can be expressed in
terms of the selected universal constants as: 
\begin{equation}
\label{1} 
\begin{array}{l}
l_p= 
\sqrt{\frac{G\hbar }{c^3}}\approx 10^{-34.7915}\text{ }m \\ t_p= 
\sqrt{\frac{G\hbar }{c^5}}\approx 10^{-43.2684}\text{ }s\qquad \\ m_p= 
\sqrt{\frac{\hbar c}G}\approx 10^{-7.6622}\text{ }kg \\ \theta _p=\sqrt{
\frac{\hbar c^5}{k_B^2G}}\approx 10^{32.1514}K 
\end{array}
\end{equation}
The meaning of these quantities is not yet quite settled, but there is no
doubt about their importance. As Planck himself states:\\

{\em '' these quantities maintain their natural significance as long as the
laws of gravitation, of propagations of light in vacuum and of both first
and second laws of thermodynamics, remain valid. They must consequently
always turn up the same, even when measured by most different inteligent
beings with the most different methods ''}\\

A more modern insight allows for $l_p$ two posible views: either that the
nature of physical phenomena changes radically for lenghs lower than $l_p$%
{\em \ ,} or that {\em \ }$l_p$ is the lowest limit of lenght what amounts
to state that we cannot distinguish between points whose distance is lowest
than $l_p.$ Planck`s time, $t_p$, is supposed to replace the original
singularity in the very early universe(see $\left[ 3\right] )$. Planck`s
mass, $m_p,$ seems to represent a maximon, i.e. the final mass of a mini
black hole submitted to Hawking`s radiation(see $\left[ M\right] )$.
Planck`s temperature, $\Theta _p$ , can be interpreted as the temperature in
the very early universe (big-bang) with pressure predominance.

Currently Planck`s charge is calculated as a relation among the constants $%
G,c,\hbar $ what corresponds to a mutilated base (see $\left[ 6\right] )$ $%
\left\{ L,M,T\right\} $ and the expresion 
\begin{equation}
\label{2}q_p=\sqrt{\hbar c} 
\end{equation}
is unavoidably obtained. That result is in our view, untenable, since the
dimensions of electrical charge cannot be reducible to mechanical ones. This
is one of the reasons that we have considered to propose an enlargement of
Plack`s system. A further enlargement derives from the fact that force and
rate of change of linear momentum are different quantities. These features
justify, in our view, the enlargement of Planck System that follows.

\section{\bf \ Enlargement of the System.}

Our first step is adding to the constans sellected by Planck $\left(
G,c,\hbar ,k_B\right) $ the {\bf dynamical constant $\Gamma $} , already
considered by Bridgman in another context (see $\left[ 1\right] )$ 
\begin{equation}
\overrightarrow{F}=\Gamma \frac{d(m\overrightarrow{v})}{dt} 
\end{equation}
$\Gamma $ relates inertial force to the rate of change of linear moment, and
parallels the expresion for gravitational force $F=G\frac{mm^{\prime }}{r^2}%
, $ that justifies the presence of $G$. In our opinion the main reason for
the general acceptance of $G$ as a universal constant, derives from the fact
that making use of conventional units , its numerical value is very small.
This is not ''a priori'' the case with $\Gamma $ that has been, in our view
improperly excluded. Its inclusion increases the $\left\{ L,M,T\right\} $
base to $\left\{ L,M,T,F\right\} $ where $F$ stands for dimensions of force.
In this base the proposed universal constant, $\Gamma ,$ acquire the
dimensions 
\begin{equation}
\label{4}\left[ \Gamma \right] =FM^{-1}L^{-1}T^2 
\end{equation}
In this enlarged system the units of length, time, mass and temperature
stated in (1) become: 
\begin{equation}
\label{5} 
\begin{array}{l}
l_p^{\prime }= 
\sqrt{\frac{G\hbar }{\Gamma c^3}}=\Gamma ^{-\frac 12}\cdot l_p \\ 
t_p^{\prime }= 
\sqrt{\frac{G\hbar }{\Gamma c^5}}=\Gamma ^{-\frac 12}\cdot t_p \\ 
m_p^{\prime }= 
\sqrt{\frac{\hbar c\Gamma }G}=\Gamma ^{\frac 12}\cdot m_p \\ \theta
_p^{\prime }=\sqrt{\frac{\hbar c^5\Gamma }G}=\Gamma ^{\frac 12}\cdot \theta
_p 
\end{array}
\end{equation}
In order to give an adequate treatment to electric force we must replace the
old Culomb`s law in a vacuum 
\begin{equation}
\label{6}F=\frac{qq^{\prime }}{r^2}\qquad by\qquad F=\frac{qq^{\prime }}{%
\epsilon _0^{\prime }r^2} 
\end{equation}
Where $\epsilon _0^{\prime }=4\pi \epsilon _0$ ,being $\epsilon _0$ the {\bf %
permittivity of the vacuum }in a rationalized form. Dimensionally $\left[
\epsilon _0^{^{\prime }-1}q_p^2\right] =L^2F$ . Eq. $\left( 6\right) $
strictly applies only to stationary charges, but, dimensionally, nothing
changes if, invoking the special theory of relativity, we replace it by 
\begin{equation}
\label{7}\overrightarrow{f}=\frac 1{\epsilon _0^{\prime }c^2}\frac{%
(q^{\prime }\overrightarrow{v})\wedge (q\overrightarrow{v})}{r^2} 
\end{equation}
The expression for Planck`s charge, $q_p$ , can be obtained through the
following dimensional matrix: 
\begin{equation}
\label{8}\left( 
\begin{array}{rrrrrr}
& G & \Gamma & \hbar & c & \epsilon _0^{\prime -1}q_p^2 \\ 
L & 2 & -1 & 2 & 1 & 2 \\ 
M & -2 & -1 & 1 & 0 & 0 \\ 
T & 0 & 2 & -1 & -1 & 0 \\ 
F & 1 & 1 & 0 & 0 & 1 
\end{array}
\right) 
\end{equation}
giving a single non-dimensional product $\pi _1$ that leads to 
\begin{equation}
\label{9}q_p=\sqrt{\Gamma \epsilon _0^{\prime }\hbar c} 
\end{equation}
Since $\Gamma $ has not yet been numerically defined, we are free to make
the best selection for the value of $q_p$ . In our view the most
advantageous and simple hypothesis is to assimilate Planck`s charge to the
electron charge $\left( e^{-}=10^{-18.795289}C\right) .$ Then: 
\begin{equation}
\label{10}\qquad \Gamma =\frac{e^2}{\epsilon _0^{\prime }c\hbar }=\frac{e^2}{%
2ch\epsilon _0}\equiv \alpha \text{ \quad {\bf ( fine-structure const. )}} 
\end{equation}
This completes the enlargement.

\section{\bf Consequences of the proposed enlargement.}

\subsection{The constants to be considered as units.}

The constants to be considered as units in the enlarged system are admitedly 
$\left\{ G,c,\hbar ,k_B\right\} $ and now also $\left\{ \Gamma =\alpha
,e,\epsilon _0,\mu _0^{\prime }=\frac{\mu _0}{4\pi }\right\} .$ All of them
will acquiere the value unity as results from $\left( 10\right) $ and from
the relation 
\begin{equation}
\label{11}\epsilon _0\mu _0=\frac 1{c^2}=\epsilon _0^{\prime }\mu _0^{\prime
} 
\end{equation}

\subsection{Numerical values.}

After eliminations of $\Gamma =\alpha $ between $\left( 10\right) $ and $%
\left( 5\right) $ taking account of $\left( 1\right) $ we obtain 
\begin{equation}
\label{12} 
\begin{array}{l}
l_p^{\prime }= 
\sqrt{\frac{G\hbar }{\Gamma c^3}}=\sqrt{\frac{G\hbar ^2\epsilon _0^{\prime } 
}{e^2c^2}}\approx 10^{-33.72313}\text{ }m \\ t_p^{\prime }= 
\sqrt{\frac{G\hbar }{\Gamma c^5}}=\sqrt{\frac{G\hbar ^2\epsilon _0^{\prime } 
}{e^2c^4}}\approx 10^{-42.19995}s \\ m_p^{\prime }= 
\sqrt{\frac{\hbar c\Gamma }G}=\sqrt{\frac{e^2}{G\epsilon _0^{\prime }}}%
\approx 10^{-8.730614}\text{ }kg \\ k_B\theta _p^{\prime }\ = 
\sqrt{\frac{\hbar c^5\Gamma }G}=\sqrt{\frac{e^2c^4}{G\epsilon _0^{\prime }}}%
\approx 10^{31.08294}K \\ E_p^{\prime }=m_p^{\prime }c^2=\frac{ec^2}{\sqrt{%
G\epsilon _0^{\prime }}}\approx 10^{8.22303}J 
\end{array}
\end{equation}
All of this results have been calculated in the International System.

\subsection{Temperature in a Friedman Robertson-Walker model.}

Using E.P.S. the known expresion of temperature in a FRW with pressure
predominance is easily obtained: 
\begin{equation}
\label{13}\theta _R^{\prime }\propto \theta _p^{\prime }\ \left( \frac{%
t_p^{\prime }}t\right) ^{\frac 12}=\left( \frac{c^4\hbar ^2e^2}{G\epsilon
_0^{\prime }k_B^4}\right) ^{\frac 14}\cdot t^{-\frac 12}\approx
10^{9.98297}t^{-\frac 12} 
\end{equation}
from $(5)$ follows the relation 
\begin{equation}
\label{14}\theta _R\propto \theta _p\ \left( \frac{t_p}t\right) ^{\frac
12}=\left( \Gamma \right) ^{-\frac 14}\theta _R^{\prime } 
\end{equation}

\subsection{Homogeneity and isotropy.}

Commonly Planck volumen $\tau _p$ is taken as $\tau _p=l_p^3$ . At the
begining of this paper we considered $l_p$ as the lowest limit of length
what amounts to state that we cannot distinguish between points whose
distance is lower than $l_p.$ Consequently we ought to consider the volume
of a sphere with diameter $l_p$ as the lowest limit of Planck`s volume. Then 
$\tau _p\geq \frac \pi 6l_p^3$ and we can write $\tau _p^{\prime }\geq
bl_p^{\prime 3}$ where $b\geq \frac \pi 6$ remains to be calculated.

We define the Planck`s energy density, in the E.P.S. as: 
\begin{equation}
\label{15}\rho _p^{\prime }=\frac{k_B\theta _p^{\prime }}{b\cdot \left(
l_p^{\prime }\right) ^3}=\frac{c^5e^4}{b\hbar ^3\left( \epsilon _0^{\prime
}G\right) ^2}\qquad \qquad \frac \pi 6\leq b 
\end{equation}
Admitedly (see$\left[ 4\right] $) in the very early universe the following
expresion ought to be applied 
\begin{equation}
\label{16}\rho _R=\frac{\pi ^2}{30}g(T)\frac{k_B^4}{\hbar ^3c^3}\left(
\theta _p\right) ^4 
\end{equation}
where $g(T)$ represents spin degenerancy factor.

In the E.P.S. the last expresion becomes: 
\begin{equation}
\label{17}\rho _R^{\prime }=\frac{\pi ^2}{30}\frac{e^4c^5}{\epsilon
_0^{\prime 2}\hbar ^3G^2}g(T) 
\end{equation}
In the case of thermal radiation, the particles will be only photons and
then $g(T)=2$ . In such a case from $(17)$ and $\left( 15\right) $ follows
the ratio: 
\begin{equation}
\label{18}\frac{\rho _R^{\prime }}{\rho _p^{\prime }}=\frac{\pi ^2b}{15} 
\end{equation}
and the value $b=\frac{15}{\pi ^2}>\frac \pi 6$ means initial isotropy and
homogeneity. This correspond to define the volume of Planck by 
\begin{equation}
\label{19}\tau _p=\frac{15}{\pi ^2}\left( l_p^{\prime }\right)
^3=10^{0.18179}\left( l_p^{\prime }\right) ^3 
\end{equation}

\section{\bf Remark}

In this paper we have proposed an approach to a fruitful enlargement of
Planck System. However we want to remark that most of the results may be
quantitativelly altered. We have selected the set of universal constant most
generally considered as constituting Planck System $\left( G,c,\hbar
,k_B\right) .$ But it migth prove advisable to select $h$ instead of $\hbar
, $ and/or $a=\frac{\pi ^2}{15}$ $\frac{k_B^4}{\hbar ^3c^3}$ (Boltzman
constant) instead of $k_B$ , to give only two examples among many possibles.
The best fitting of the numerical results to their most probable values is
perhaps the most suitable guideline.\

\end{document}